\documentclass[aps, prb, twocolumn,amsmath,amssymb,floatfix, reprint]{revtex4-2}
\usepackage{graphicx}
\usepackage{physics}
\usepackage{times}
\usepackage{dcolumn}
\usepackage{hyperref} % For hyperlinks in the PDF
\hypersetup{colorlinks, allcolors=blue}
\begin{document}

\title{Quantum fluctuations in the order parameter of quantum skyrmion crystals}

\author{Kristian M{\ae}land} 
\affiliation{\mbox{Center for Quantum Spintronics, Department of Physics, Norwegian University of Science and Technology, NO-7491 Trondheim, Norway}} 
\author{Asle Sudb{\o}}
\email[Corresponding author: ]{asle.sudbo@ntnu.no}
\affiliation{\mbox{Center for Quantum Spintronics, Department of Physics, Norwegian University of Science and Technology, NO-7491 Trondheim, Norway}} 
    
% \affiliation{\mbox{Center for Quantum Spintronics, Department of Physics, Norwegian University of Science and Technology,}\\NO-7491 Trondheim, Norway} 
% \affiliation{\mbox{Center for Quantum Spintronics, Department of Physics, Norwegian University of Science and Technology, NO-7491 Trondheim, Norway}} 

%\date{\today} 

\begin{abstract}
Traditionally, skyrmions are treated as continuum magnetic textures of classical spins with a well-defined topological skyrmion number. Owing to their topological protection, skyrmions have attracted great interest as building blocks in future memory technology. Smaller skyrmions offer greater memory density, however, quantum effects are not negligible for skyrmions with sizes of just a few lattice constants. In this paper we study quantum fluctuations around dense skyrmion crystal ground states, and focus on the utility of a discretized order parameter for quantum skyrmions. The order parameter is found to be a useful indicator of the existence of quantum skyrmions, and captures the quantum phase transition between two distinct skyrmion phases.
\end{abstract}

\maketitle

%------------------------------------------INTRODUCTION--------------------------------------------------------------------------
\section{Introduction}
%Skyrmions are traditionally thought of as nontrivial textures of classical spins with a well-defined topological index, also referred to as a winding number or topological charge,  involving a two-dimensional integral over continuum space \cite{nagaosaRev, KlauiRev2016}. 
Skyrmions are traditionally thought of as nontrivial textures of classical spins with a well-defined topological skyrmion number, also referred to as a winding number or topological charge. The topological skyrmion number involves a two-dimensional integral over continuum space \cite{nagaosaRev, KlauiRev2016}. 
As such, skyrmions are conventionally assumed to be quite large objects involving many spins. From a fundamental physics point of view, it is of interest to see to what extent one can characterize such nontrivial spin textures involving far fewer spins, also taking into account the intrinsic quantum nature of individual spins.      

Skyrmions have attracted significant interest in recent years due to promising future applications in magnetic memory technology \cite{nagaosaRev, KlauiRev2016, JonietzCurrentControlSK,skyrmionroadmap,  RacetrackTomasello, RactrackFert, DMIguideSmallJ, ControlSmallJ,  ExchangeFreeJ0,smallSkRW, hsuWriteDeleteSk, yu2017SkRead}. Compared to conventional memory technology based on regions of uniform magnetization, the existence or nonexistence of a skyrmion can be used as a more stable bit due to its topological protection \cite{nagaosaRev, KlauiRev2016, je2020topostab}. 
%Methods to write, delete and detect single skyrmions have been proposed \cite{ExchangeFreeJ0, skyrmionroadmap} and realized \cite{smallSkRW, hsuWriteDeleteSk, yu2017SkRead, skyrmionroadmap}. Furthermore, it has been found that skyrmion dynamics driven by currents require ultralow current densities \cite{JonietzCurrentControlSK} compared to domain wall motion \cite{nagaosaRev}. Therefore, skyrmions are considered as promising candidates for future racetrack memory devices \cite{KlauiRev2016, Parkinracetrackdomain, skyrmionroadmap, RacetrackTomasello, RactrackFert}. Skyrmions may also find applications as information carriers in spin logic gates \cite{zhangSkAND, KlauiRev2016} and as qubits in quantum computing \cite{SkQubits}. 
Furthermore, skyrmions are considered as promising candidates for future racetrack memory devices \cite{nagaosaRev, KlauiRev2016, JonietzCurrentControlSK, Parkinracetrackdomain, skyrmionroadmap, RacetrackTomasello, RactrackFert}. Skyrmions may also find applications as information carriers in spin logic gates \cite{zhangSkAND, KlauiRev2016} and as qubits in quantum computing \cite{SkQubits}. 

A periodic lattice of skyrmions, i.e., a skyrmion crystal (SkX), was first observed in the chiral magnet MnSi \cite{Muhlbauer1stSkExp}. 
%When considering skyrmions whose size is far greater than the lattice constant it is often sufficient to study the magnetization on a classical level \cite{nagaosaRev}. However, when aiming for dense memory storage it is necessary to study as small skyrmions as possible \cite{ExchangeFreeJ0}. 
When aiming for dense memory storage it is necessary to study as small skyrmions as possible \cite{ExchangeFreeJ0}. 
%In Ref.~\cite{HeinzeSkX}, a dense SkX containing nanometer sized skyrmions was observed in a monolayer of Fe on the triangular Ir(111) surface. 
Using spin-polarized scanning tunneling microscopy, a dense SkX containing nanometer-sized skyrmions was observed in a monolayer of Fe on the triangular Ir(111) surface in Ref.~\cite{HeinzeSkX}.
For skyrmions whose size is only a few lattice constants, a quantum description is necessary \cite{Sotnikov, Lohani, gauyacq2019, Ochoa}. We therefore refer to these skyrmions as quantum skyrmions. The discrete nature of quantum skyrmions casts doubt on their topological protection, since topological arguments rely on continuum descriptions. However, it has been shown that skyrmions are topologically stable in a real, discrete system \cite{je2020topostab}.

Studying the spin waves and the quantum fluctuations is important to understand the dynamics of quantum skyrmions \cite{RoldanMolina, RoldanMolina2016, dosSantosPRB, DiazPRL, DiazPRR}. In this paper, we obtain the spin waves on top of a SkX ground state (GS) reminiscent of that observed in Ref.~\cite{HeinzeSkX}, as well as a distinct SkX with the same periodicity. %Our main focus is on the quantum fluctuations of the discretized, chiral order parameter (OP) \cite{Sotnikov, Lohani, BERGL}, essentially asking how we can best classify the existence of a quantum skyrmion when the continuum and inherently classical, topological skyrmion number is not well defined \cite{nagaosaRev, Sotnikov}. 
Our main focus is on the quantum fluctuations of the discretized order parameter (OP) for quantum skyrmions \cite{Sotnikov, Lohani, BERGL}. 
Inspired by the work of Ref.~\cite{Sotnikov} we are considering how to classify the existence of quantum skyrmions when the topological skyrmion number is not well defined \cite{nagaosaRev, Sotnikov}. 
%In Ref.~\cite{Sotnikov}, exact diagonalization was used on a 19-site cluster and the quantum scalar chirality was introduced to probe the existence of quantum skyrmions. We build on this by instead considering a truly periodic quantum SkX in zero external magnetic field. While the Holstein-Primakoff approach relies on quantum fluctuations around a classical GS, it permits a calculation of a quantum skyrmion OP that is more closely related to the topological skyrmion number.
In Ref.~\cite{Sotnikov}, exact diagonalization was used on a 19-site cluster. We use the Holstein-Primakoff (HP) \cite{HP-PhysRev.58.1098} approach to consider quantum fluctuations around a classical GS. This involves certain approximations whose validity will be commented on. While exact diagonalization involves an exact treatment of the quantum nature of a limited number of spins, the HP approach allows us to treat truly periodic lattices of quantum skyrmions.

The paper is organized as follows. Sec.~\ref{sec:model} introduces the model and the SkX GSs, while details of the method to obtain the GSs are reserved for Appendix \ref{app:GS}. The magnon description is introduced in Sec.~\ref{sec:magnon} and the magnon energy spectrum is discussed in Sec.~\ref{sec:spectrum}, whereas details of the diagonalization procedure are given in Appendix \ref{app:colpa}. The main results of the paper regarding the quantum skyrmion OP are given in Sec.~\ref{sec:OP}. We also consider the sublattice magnetizations in Sec.~\ref{sec:fluct} before providing our conclusions in Sec.~\ref{sec:Con}. In Appendix \ref{app:Thermal} we show the effect of thermal fluctuations on the quantum skyrmion OP.

%----------------------------------------------------MODEL-----------------------------------------------------------------
\section{Model and ground states} \label{sec:model}
%Hamiltonian, origin from Heinze paper, DMI, four-spin interaction

Inspired by the $1~\text{nm}\cross 1~\text{nm}$ skyrmions observed in Ref.~\cite{HeinzeSkX}, we adopt the following Hamiltonian,
\begin{equation}
\label{eq:H}
    H = H_{\text{ex}} + H_{\text{DM}} + H_{\text{A}} + H_{4},
\end{equation}
where
% \begingroup
% \allowdisplaybreaks
% \begin{align}
%     H_{\text{ex}}  =& -J\sum_{\langle ij \rangle} \boldsymbol{S}_{i}\cdot \boldsymbol{S}_{j},  \\
%     H_{\text{DM}} =& \sum_{\langle ij \rangle} \boldsymbol{D}_{ij} \cdot (\boldsymbol{S}_i \cross \boldsymbol{S}_j),   \\
%     H_{\text{A}} =& - K\sum_i S_{iz}^2,    \\
%     H_4 =& U\sum_{ijkl}^\diamond \big[(\boldsymbol{S}_i \cdot \boldsymbol{S}_j)(\boldsymbol{S}_k \cdot \boldsymbol{S}_l) + (\boldsymbol{S}_i \cdot \boldsymbol{S}_l)(\boldsymbol{S}_j \cdot \boldsymbol{S}_k) \nonumber \\
%     &\mbox{\qquad\quad}-(\boldsymbol{S}_i \cdot \boldsymbol{S}_k)(\boldsymbol{S}_j \cdot \boldsymbol{S}_l)\big].
% \end{align}
% \endgroup
\begin{equation}
    H_{\text{ex}}  = -J\sum_{\langle ij \rangle} \boldsymbol{S}_{i}\cdot \boldsymbol{S}_{j},
\end{equation}
\begin{equation}
    H_{\text{DM}} = \sum_{\langle ij \rangle} \boldsymbol{D}_{ij} \cdot (\boldsymbol{S}_i \cross \boldsymbol{S}_j),
\end{equation}
\begin{equation}
        H_{\text{A}} = - K\sum_i S_{iz}^2, 
\end{equation}
\begin{align}
    H_4 =& U\sum_{ijkl}^\diamond \big[(\boldsymbol{S}_i \cdot \boldsymbol{S}_j)(\boldsymbol{S}_k \cdot \boldsymbol{S}_l) + (\boldsymbol{S}_i \cdot \boldsymbol{S}_l)(\boldsymbol{S}_j \cdot \boldsymbol{S}_k) \nonumber \\
    &\mbox{\qquad\quad}-(\boldsymbol{S}_i \cdot \boldsymbol{S}_k)(\boldsymbol{S}_j \cdot \boldsymbol{S}_l)\big].
\end{align}
Here, $\boldsymbol{S}_i$, with magnitude $S$, is the spin operator at lattice site $i$ located at $\boldsymbol{r}_i$. We consider a triangular lattice in the $xy$ plane. 
%The description is limited to nearest-neighbor ferromagnetic exchange interaction, indicated by $\langle ij \rangle$. 
The ferromagnetic exchange interaction is limited to nearest neighbors, indicated by $\langle ij \rangle$.
The Dzyaloshinskii-Moriya interaction (DMI) \cite{DZYALOSHINSKY, Moriya} favors noncollinear magnetization, as opposed to the exchange interaction. DMI is possible for the triangular lattice under consideration because it is assumed to be on a surface, and so inversion symmetry is broken in the $z$ direction \cite{Moriya, HeinzeSkX, KlauiRev2016}. The DMI vector is set to $\boldsymbol{D}_{ij} = D \hat{r}_{ij} \cross \hat{z}$, where $\hat{r}_{ij}$ is the unit vector connecting nearest neighbors $i$ and $j$, and $D>0$ is the strength of the DMI \cite{HeinzeSkX}. We also consider a single-ion easy-axis anisotropy as well as the four-spin interaction $H_4$, where the sum is limited to sites $i,j,k,l$ that make counterclockwise diamonds of minimal area, indicated by $\diamond$ \cite{HeinzeSkX, H4PRB}. The four-spin interaction makes both ferromagnetic and helical states unfavorable, and is instrumental in stabilizing the nanometer sized SkX \cite{HeinzeSkX}. The reduced Planck's constant and the lattice constant are set to $1$ throughout this paper, $\hbar = a = 1$.

\begin{figure*}
    \centering
    \begin{minipage}{.66\textwidth}
      \includegraphics[width=\linewidth]{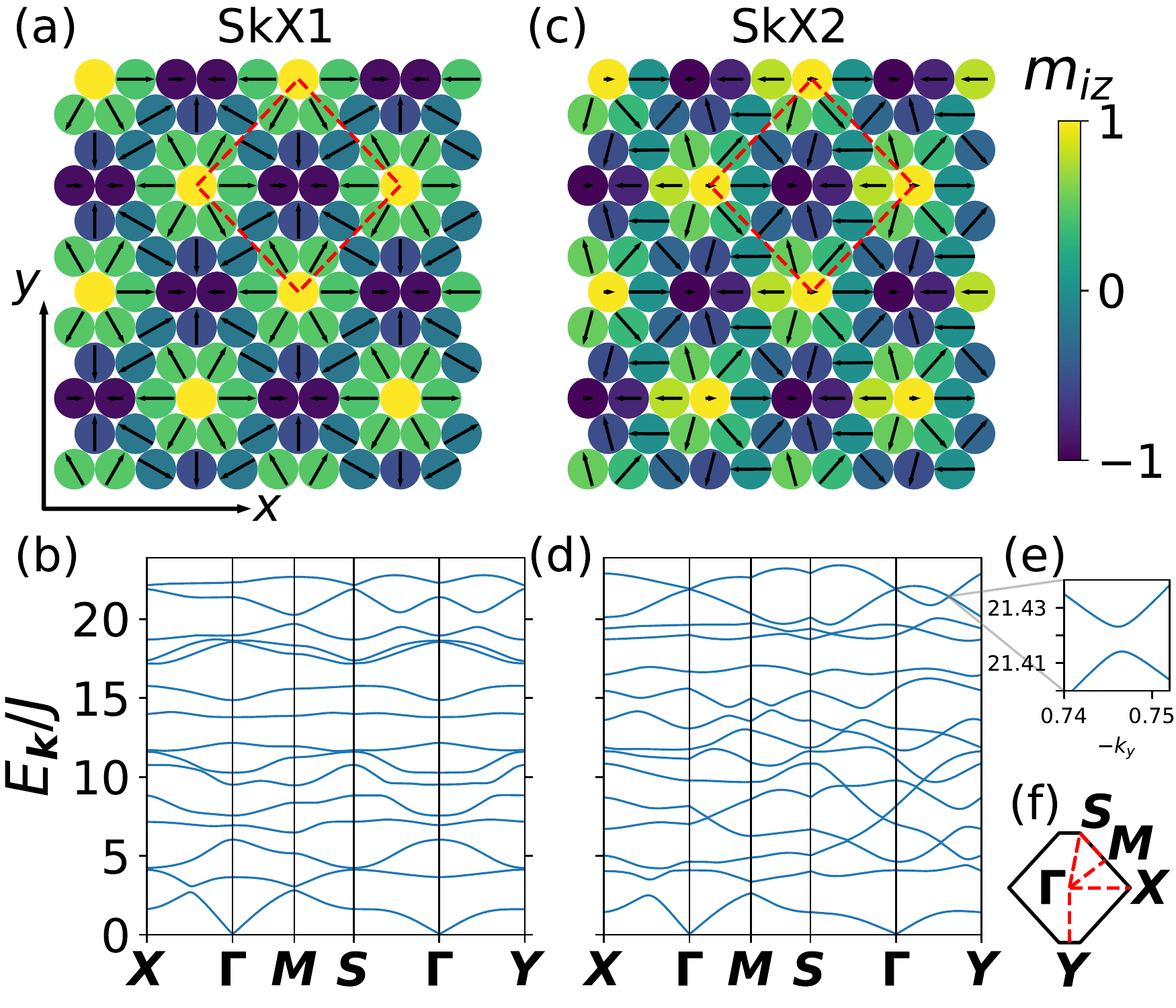}
    \end{minipage}%
    \begin{minipage}{.34\textwidth}
      \includegraphics[width=\linewidth]{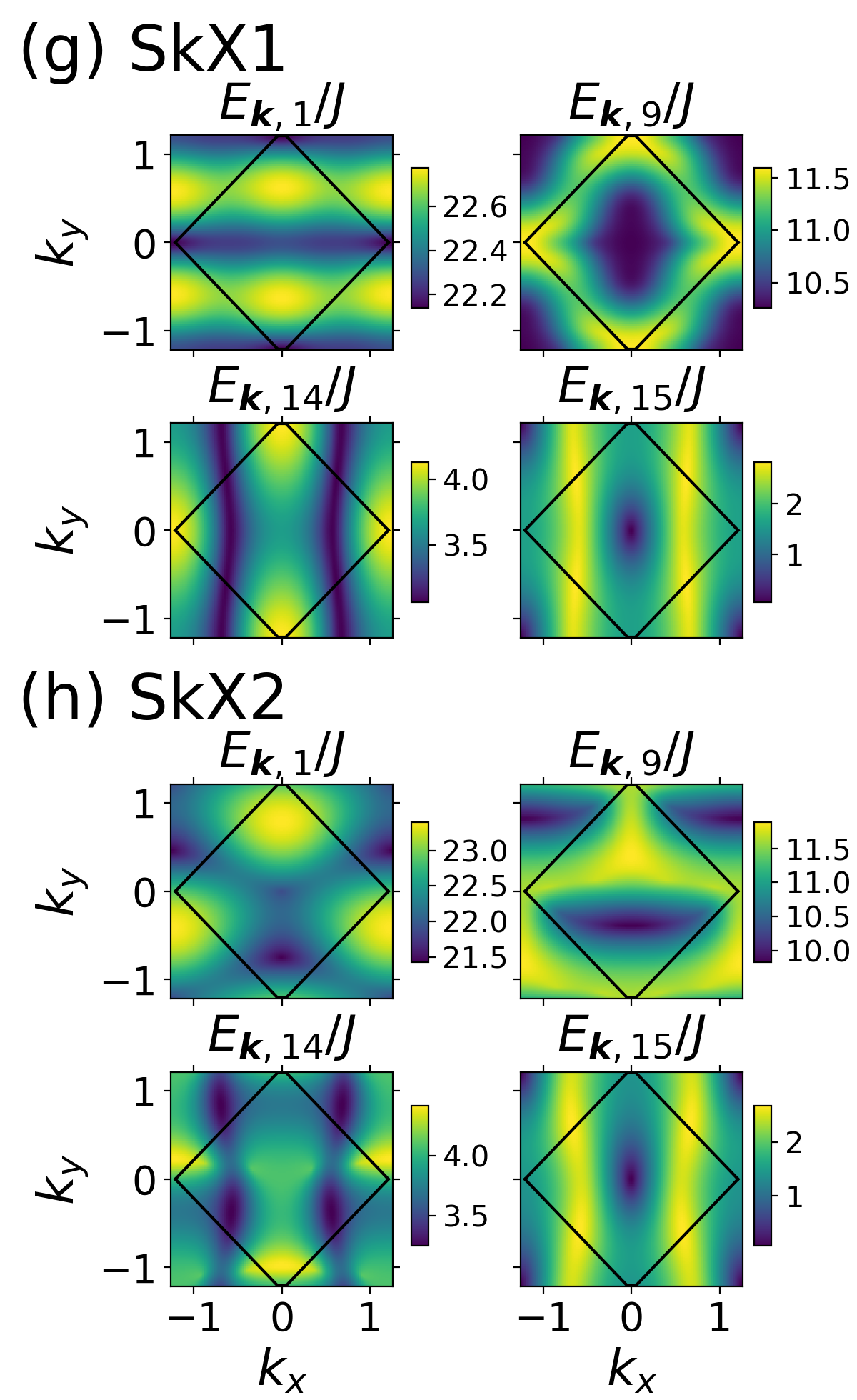}
    \end{minipage}%
    \caption{(a,c) Skyrmion crystal (SkX) ground state shown on a $10\cross12$ lattice with periodic boundary conditions. Colors indicate the $z$ component of the spins, while the arrows show their projection on the $xy$ plane. The primitive cell of the sublattices is shown in dashed red. (a) shows the state SkX1, while (c) shows the state SkX2. The parameters are $D/J = 2.16$, $U/J = 0.35$, $S = 1$, $K/J = 0.518$ for SkX1, and $K/J = 0.519$ for SkX2. (b) and (d) show the magnon spectrum $E_{\boldsymbol{k}}/J$ along the path in the first Brillouin zone (1BZ) sketched in (f), for SkX1 and SkX2, respectively. For SkX1 at $K/J = 0.518$ and SkX2 at $K/J = 0.519$ none of the bands cross at any $\boldsymbol{k}$. An example of an avoided crossing in the spectrum of SkX2 is shown in (e), where $E_{\boldsymbol{k},1}$ and $E_{\boldsymbol{k},2}$ are plotted along $-k_y$ with $k_x = 0$. (g) and (h) show the magnon energies $E_{\boldsymbol{k},1}, E_{\boldsymbol{k},9}, E_{\boldsymbol{k},14},$ and $E_{\boldsymbol{k},15}$ for SkX1 and SkX2, respectively. The 1BZ is indicated with black lines.}
    \label{fig:GS_spectrum}
\end{figure*}

By tuning the parameters to $D/J = 2.16$ and $US^2/J = 0.35$ the GS is a SkX with a periodicity of five lattice sites in the $x$ direction and six chains in the $y$ direction, at least in the region $0 \leq K/J \leq 0.8$ of easy-axis anisotropy. This is the same periodicity as the best commensurate approximation to the SkX observed in Ref.~\cite{HeinzeSkX}. 
%By applying self-consistent iteration \cite{dosSantosPRB}, the GS is as shown in Fig.~\ref{fig:GS_spectrum}(a,c). 
We find a quantum phase transition (QPT) between two distinct SkX states with the same periodicity, named SkX1 and SkX2. The QPT occurs at a transition point $K = K_t$, where $0.518 < K_t/J < 0.519$, which is determined by the classical GS energy of the two phases. 
% The QPT occurs at a critical easy-axis anisotropy  $0.518 < K_c/J < 0.519$ \footnote{We use the term {\textit{critical easy-axis anisotropy}} $K_c$, even though the quantum phase transition from SkX1 to SkX2 appears to be a discontinuous one, and not a quantum critical phenomenon.}, which is determined by the classical GS energy of the two phases. 
SkX1 is shown for $K/J = 0.518$ in Fig.~\ref{fig:GS_spectrum}(a), while SkX2 is shown for $K/J = 0.519$ in Fig.~\ref{fig:GS_spectrum}(c). They are separated by the location of the center of the skyrmion, which is placed exactly at a specific lattice site in SkX1, but has moved approximately one quarter lattice constant to the left of a specific lattice site in SkX2. Ref.~\cite{Sotnikov} also found that the center of the skyrmion relocated when tuning a parameter in the Hamiltonian, there in response to an increasing magnetic field. Here, the center relocates due to stronger easy-axis anisotropy, since the DMI and easy-axis anisotropy terms favor SkX2, while the exchange and four-spin interaction terms favor SkX1. 

Each phase appears very similar at all considered $K$, except that the absolute value of the $z$ component of all spins increases with increasing $K$. For both phases, the magnetic unit cell consists of 15 lattice sites hosting one skyrmion, see Fig.~\ref{fig:fluct}(a). This is nearing the lower limit of skyrmion size on a triangular lattice \cite{Sotnikov, ExchangeFreeJ0, gauyacq2019}. The Hamiltonian in Eq.~\eqref{eq:H} is symmetric under spin inversion, i.e., time-reversal symmetric, so states with all spins flipped compared to those in Fig.~\ref{fig:GS_spectrum}(a,c) are degenerate in energy. Additionally, SkX2 is degenerate with a state where the center relocates by the same length to the right. If these SkXs were created by cooling down an unordered magnetic state in a suitable material, domain walls would probably appear between the degenerate states \cite{HeinzeSkX, HeinzeSkXMC}.

A sizable value of $D/J$ is needed in our model to stabilize the quantum skyrmions. DMI originates from spin-orbit coupling (SOC), which is a relativistic effect, and therefore $D$ is usually much smaller than $J$ in real materials \cite{Moriya, Dsmall_vdWTHE, KlauiRev2016}. However, methods of tuning the value of $J$ using external lasers \cite{DMIguideSmallJ, ControlSmallJ}, even to the point of an exchange free magnet  \cite{ExchangeFreeJ0}, have recently been put forth. Therefore, studying materials with $D>J$ in the laboratory has become feasible. Additionally, in the material in Ref.~\cite{HeinzeSkX}, the large nuclear number of Ir leads to a strong SOC and therefore enhances the DMI. Moreover, the nearest-neighbor exchange interaction is unusually weak due to the strong hybridization of Fe and Ir, which is also why the four-spin interaction is not negligible \cite{HeinzeSkX}. While our parameter set differs from that given in Ref.~\cite{HeinzeSkX}, we are modeling a similar material, namely a ferromagnetic monolayer on the surface of a heavy metal which gives rise to strong SOC and a weak nearest-neighbor exchange interaction \cite{HeinzeSkX, KlauiRev2016}. However, achieving $D/J = 2.16$ in a real material still might require tuning $J$ \cite{ControlSmallJ} to a lower value. 

We also present results as functions of the easy-axis anisotropy. Ordinarily, this is considered to be a constant value due to the magnetocrystalline anisotropy in a given material \cite{HeinzeSkX, RoldanMolina, RoldanMolina2016, dosSantosPRB}. Therefore one would need to look at different materials in order to observe the changes in the results. However, a method to tune the value of $K$ in a material by applying mechanical strain, even managing to change its sign and so change its nature to an easy-plane anisotropy, has recently been discovered \cite{TuneK_PRB, TuneK}. Motivated by this, we will discuss a tunable easy-axis anisotropy $K$. Since the phase transition between SkX1 and SkX2 occurs at zero temperature by tuning a parameter in the Hamiltonian, we identify it as a QPT. 

%-----------------------------MAGNON DESCRIPITON-----------------------
\section{Magnon description} \label{sec:magnon}
%Apply Haraldsen-Fishman spin rotation technique to Hamiltonian based on our GS. Fourier transform and write on matrix form.
% Ref.~\cite{HProtation_2009} introduces a method of performing the Holstein-Primakoff (HP) transformation \cite{HP-PhysRev.58.1098} in noncollinear magnetic ground states. 
Ref.~\cite{HProtation_2009} introduces a method of performing the HP transformation in noncollinear magnetic ground states. 
For each lattice site the approach introduces rotated coordinates through an orthonormal frame $\{\hat{e}_1^i, \hat{e}_2^i, \hat{e}_3^i\}$ with $\hat{e}_3^i = \boldsymbol{m}_i$. Here, $\boldsymbol{m}_i = (\sin\theta_i\cos\phi_i, \sin\theta_i\sin\phi_i, \cos\theta_i)$ is a unit vector along the spin at lattice site $i$ in the classical GS. Furthermore, $\hat{e}_1^i = (\cos\theta_i\cos\phi_i, \cos\theta_i\sin\phi_i, -\sin\theta_i)$ and $\hat{e}_2^i = (-\sin\phi_i, \cos\phi_i, 0)$. The HP transformation is then introduced as $S_{i3}=\boldsymbol{S}_i \cdot \hat{e}_3^i = S - a_i^\dagger a_i$, $S_{i\pm} = \boldsymbol{S}_i \cdot \hat{e}_\pm^i $, $\hat{e}_\pm^i = \hat{e}_1^i \pm i\hat{e}_2^i$, $S_{i+} = \sqrt{2S}a_i,$ and $S_{i-} = \sqrt{2S}a_i^\dagger$. Assuming small spin fluctuations, we truncate at second order in magnon operators and therefore neglect magnon-magnon interactions \cite{HProtation_2009, DiazPRL}. The assumption of small spin fluctuations should be valid at low temperatures compared to the lowest magnon energy, see Appendix \ref{app:Thermal}. 
%$S_{i+} = \sqrt{2S-a_i^\dagger a_i}~a_i,$ and $S_{i-} = a_i^\dagger\sqrt{2S-a_i^\dagger a_i}$

Let $N$ be the number of lattice sites and $N' = N/15$ the number of magnetic unit cells. The Fourier transform (FT) is introduced as $a_i = \frac{1}{\sqrt{N'}} \sum_{\boldsymbol{k}} e^{i\boldsymbol{k}\cdot \boldsymbol{r}_i} a_{\boldsymbol{k}}^{(r)}$, assuming lattice site $i$ is located on sublattice $r$. The sum over momenta $\boldsymbol{k}$ covers the first Brillouin zone (1BZ) of sublattice $r$.

The magnon-independent terms, $H_0$, correspond to the classical Hamiltonian. It can be shown that linear terms in magnon operators vanish when expanding around the GS \cite{HProtation_2009}. The quadratic terms may be written
\begin{equation}
\label{eq:H2}
    H_2 = \frac{1}{2}\sum_{\boldsymbol{k}}  \boldsymbol{a}_{\boldsymbol{k}}^\dagger M_{\boldsymbol{k}}  \boldsymbol{a}_{\boldsymbol{k}},
\end{equation}
where $\boldsymbol{a}_{\boldsymbol{k}}^\dagger = (a_{\boldsymbol{k}}^{(1)\dagger}, a_{\boldsymbol{k}}^{(2)\dagger}, \dots, a_{\boldsymbol{k}}^{(15)\dagger}, a_{-\boldsymbol{k}}^{(1)}, \dots, a_{-\boldsymbol{k}}^{(15)})$ and
\begin{equation}
\label{eq:M}
    M_{\boldsymbol{k}} = \begin{pmatrix} \eta_{\boldsymbol{k}} & \nu_{-\boldsymbol{k}}^* \\ \nu_{\boldsymbol{k}} & \eta_{-\boldsymbol{k}}^*    \end{pmatrix}.
\end{equation}
The matrix elements are expressed as
\begin{align}
    \eta_{\boldsymbol{k}}^{r,s} =& \eta_r \delta_{r,s} +Se^{i\boldsymbol{k}\cdot \boldsymbol{\delta}_{(r,s)}}\Lambda_+^{r,s},
\end{align}
\begin{align}
    \nu_{\boldsymbol{k}}^{r,s} =& \nu_r \delta_{r,s} +Se^{i\boldsymbol{k}\cdot \boldsymbol{\delta}_{(r,s)}}\Lambda_-^{r,s} ,
\end{align}
\begin{align}
    \eta_r =& 2S\sum_s [J_{(r,s)}\hat{e}_3^r \cdot \hat{e}_3^s -\boldsymbol{D}_{(r,s)} \cdot (\hat{e}_3^r \cross \hat{e}_3^s)] \nonumber \\
    &-KS[1-3(\hat{e}_3^r \cdot \hat{z})^2] \nonumber \\
    &-4S^3 \sum_{stu} U_{(r,s,t,u)}[(\hat{e}_3^r \cdot \hat{e}_3^s)(\hat{e}_3^t \cdot \hat{e}_3^u)  \nonumber \\
    & \mbox{\qquad}+ (\hat{e}_3^r \cdot \hat{e}_3^u)(\hat{e}_3^s \cdot \hat{e}_3^t) - (\hat{e}_3^r \cdot \hat{e}_3^t)(\hat{e}_3^s \cdot \hat{e}_3^u)],
\end{align}
\begin{align}
    \nu_r =& -KS(\hat{e}_1^r \cdot \hat{z})^2 ,
\end{align}
\begin{align} 
    \Lambda_\pm^{r,s} =& -J_{(r,s)}\hat{e}_{\pm}^r \cdot \hat{e}_-^s + \boldsymbol{D}_{(r,s)} \cdot (\hat{e}_{\pm}^r \cross \hat{e}_-^s) \nonumber \\
    &+2S^2\bigg(\sum_{tu} U_{(r,s,t,u)}[(\hat{e}_{\pm}^r \cdot \hat{e}_-^s)(\hat{e}_3^t \cdot \hat{e}_3^u) \nonumber \\
    &+ (\hat{e}_{\pm}^r \cdot \hat{e}_3^u)(\hat{e}_-^s \cdot \hat{e}_3^t) - (\hat{e}_{\pm}^r \cdot \hat{e}_3^t)(\hat{e}_-^s \cdot \hat{e}_3^u)] \nonumber \\
    &+\sum_{s'u} U_{(r,s',s,u)} [(\hat{e}_{\pm}^r \cdot \hat{e}_3^{s'})(\hat{e}_-^s \cdot \hat{e}_3^u) \nonumber \\
    &+ (\hat{e}_{\pm}^r \cdot \hat{e}_3^u)(\hat{e}_3^{s'} \cdot \hat{e}_-^s) - (\hat{e}_{\pm}^r \cdot \hat{e}_-^s)(\hat{e}_3^{s'} \cdot \hat{e}_3^u)] \nonumber \\
    &+\sum_{s't} U_{(r,s',t,s)} [(\hat{e}_{\pm}^r \cdot \hat{e}_3^{s'})(\hat{e}_3^t \cdot \hat{e}_-^s) \nonumber \\
    &+ (\hat{e}_{\pm}^r \cdot \hat{e}_-^s)(\hat{e}_3^{s'} \cdot \hat{e}_3^t) - (\hat{e}_{\pm}^r \cdot \hat{e}_3^t)(\hat{e}_3^{s'} \cdot \hat{e}_-^s)] \bigg).
\end{align}
Here, $\boldsymbol{\delta}_{(r,s)}$ is the shortest vector connecting a site on sublattice $r$ to a site on sublattice $s$. $J_{(r,s)} = J$ and $\boldsymbol{D}_{(r,s)} = D \boldsymbol{\delta}_{(r,s)} \cross \hat{z}$ if there exist sites $i\in r, j\in s$ such that $i$ and $j$ are nearest neighbors. Otherwise $J_{(r,s)} = 0$ and $\boldsymbol{D}_{(r,s)} = \boldsymbol{0}$. $U_{(r,s,t,u)} = U$ if there exist sites $i \in r, j \in s, k \in t, l \in u$ such that $i,j,k,l$ make a counterclockwise diamond of minimal area. Otherwise $U_{(r,s,t,u)} = 0$. Since the orthonormal frame $\{\hat{e}_1^i, \hat{e}_2^i, \hat{e}_3^i\}$ is equal for any $i \in r$ it is replaced by $\{\hat{e}_1^r, \hat{e}_2^r, \hat{e}_3^r\}$ to facilitate the FT. 

% The matrix $M_{\boldsymbol{k}}$ in Eq.~\eqref{eq:M} is positive definite in this bosonic system. Therefore, we employ the matrix generalization of the Bogoliubov transformation introduced in Ref.~\cite{COLPA} to diagonalize the Hamiltonian. 
% In Ref.~\cite{PWSWBEC}, we used the method described in Refs.~\cite{Tsallis, Xiao} because the matrix $M_{\boldsymbol{k}}$ was not positive definite in all phases. We have confirmed that all results in this paper are the same when using either diagonalization method.

%--------------------------------EXCITATION SPECTRUM-------------------------
\section{Magnon spectra} \label{sec:spectrum}
%Results for the excitation spectrum, gap i.e. no Goldstone modes \cite{DiazPRL}, some bands flat some not + no crossings \cite{Roldan-Molina}. Along path and as contour for lowest energy excitation.
Diagonalizing \cite{COLPA} and applying a bosonic commutator gives
\begin{equation}
\label{eq:H2diag}
    H_2 = \sum_{\boldsymbol{k}}\sum_{n = 1}^{15} E_{\boldsymbol{k}, n} \pqty{b_{\boldsymbol{k},n}^\dagger b_{\boldsymbol{k},n} + \frac{1}{2}}.
\end{equation}
The 15 energy bands are numbered from top to bottom, and plotted for a path in the 1BZ in Fig.~\ref{fig:GS_spectrum}(b,d). The definition of the points in the 1BZ and its precise structure are given in Appendix \ref{app:GS}. The number of bands corresponds to the number of sublattices in the GS. A selection of magnon energies are shown in the full 1BZ in Fig.~\ref{fig:GS_spectrum}(g,h). Possible experimental techniques to measure these collective excitations include spin-resolved electron-energy-loss spectroscopy \cite{dosSantosPRB}, neutron scattering, and Brillouin light scattering \cite{skyrmionroadmap}. Compared to earlier studies of SkX magnon spectra \cite{RoldanMolina, RoldanMolina2016, DiazPRL, DiazPRR, dosSantosPRB}, our results offer new insight into their behavior in zero external magnetic field.

Note the significant difference between the magnon spectrum of SkX1 in Fig.~\ref{fig:GS_spectrum}(b,g) compared to that of SkX2 in Fig.~\ref{fig:GS_spectrum}(d,h). A  striking difference is that in SkX1 the energy spectrum obeys $E(k_x, k_y) = E(-k_x, k_y) = E(k_x, -k_y)$, while in SkX2 $E(k_x, k_y) = E(-k_x, k_y) \neq E(k_x, -k_y)$. The lower symmetry of the magnon spectrum is attributed to the lower symmetry of the GS. The lower symmetry of SkX2 compared to SkX1 is also what clearly identifies it as a separate phase. Both the SkX1 GS and the SkX2 GS show a symmetry under combined real space mirror operation about the $xz$ plane and spin inversion. Meanwhile, only SkX1 shows symmetry under combined mirror operation about the $yz$ plane and spin inversion. In both states, for any nearest-neighbor spins $\boldsymbol{m}_r$ and $\boldsymbol{m}_s$ with nearest-neighbor vector $\boldsymbol{\delta}_{(r,s)}$, there exists a pair  $\boldsymbol{m}_{\overline{r}} = (m_{rx}, -m_{ry}, m_{rz})$ and $\boldsymbol{m}_{\overline{s}} = (m_{sx}, -m_{sy}, m_{sz})$ with nearest-neighbor vector $\boldsymbol{\delta}_{(\overline{r},\overline{s})} = (\delta_{(r,s)x}, -\delta_{(r,s)y})$. By inspecting the matrix elements and applying theory concerning the diagonalization method \cite{COLPA, Xiao}, this can be used to show that the matrices $M(k_x, k_y)$ and $M(-k_x, k_y)$ defined in Eq.~\eqref{eq:M} give the same energy eigenvalues.
Hence, symmetry under combined mirror operation about the $xz$ plane and spin inversion leads to mirror symmetry about the $k_y k_z$ plane for the magnon spectrum. The broken symmetry under combined mirror operation about the $yz$ plane and spin inversion in SkX2 leads to a broken mirror symmetry about the $k_x k_z$ plane for its magnon spectrum, unlike SkX1.
%Meanwhile, it is only true in the SkX1 state that for every set of nearest neighbors $\boldsymbol{m}_r$ and $\boldsymbol{m}_s$ with nearest neighbor vector $\boldsymbol{\delta}_{(r,s)}$ there exists a nearest neighbor set $\boldsymbol{m}_{\Tilde{r}} = (-m_{rx}, m_{ry}, m_{rz})$ and $\boldsymbol{m}_{\Tilde{s}} = (-m_{sx}, m_{sy}, m_{sz})$ with nearest neighbor vector $\boldsymbol{\delta}_{(\Tilde{r},\Tilde{s})} = (-\delta_{(r,s)x}, \delta_{(r,s)y})$. Therefore, it is only in the SkX1 state that $M(k_x, k_y)$ and $M(k_x, -k_y)$ give the same energy eigenvalues.

For SkX1, the magnon gap at the $\boldsymbol{\Gamma}$ point, $\Delta \equiv \min_{\boldsymbol{k}} E_{\boldsymbol{k}}$, is caused by a combination of the easy-axis anisotropy and the DMI. $H_{\text{ex}}$ and $H_{4}$ involve inner products of spins, and are independent of a global rotation of all spins. Therefore, they should not cause a gap at the $\boldsymbol{\Gamma}$ point. $H_{\text{A}}$ and $H_{\text{DMI}}$ protect the GS from a global rotation of all spins, i.e., they break continuous rotational symmetry \cite{DiazPRL}, and therefore they both contribute to the gap at the $\boldsymbol{\Gamma}$ point. Fig.~\ref{fig:fluct}(b) shows how the gap changes when $K$ is tuned. Unlike ferro- and antiferromagnets, the gap decreases when the easy-axis anisotropy is increased in the SkX1 phase. For SkX2, the magnon gap has moved slightly away from the $\boldsymbol{\Gamma}$ point toward positive $k_y$. We attribute this to the lower symmetry of the SkX2 state. Interestingly, the gap becomes an increasing function of $K$ in the SkX2 phase. This further illustrates the difference between the SkX1 and SkX2 phases, and lies at the heart of why the system transitions to the new phase at strong easy-axis anisotropy, see Appendix \ref{app:GS}. Apart from the dependence of the magnon gap on $K$ and a finite number of $K$ values where two bands cross \cite{upcoming}, the magnon spectrum in each phase is very similar at all considered values of $K$.

%---------------------------------Skyrmion OP----------------------------------
\section{Quantum skyrmion order parameter}\label{sec:OP}
% The classical, continuum, topological skyrmion number counts the number of times the magnetization wraps the unit sphere \cite{nagaosaRev}.
The topological skyrmion number counts the number of times the magnetization wraps the unit sphere \cite{nagaosaRev}. It relies on a continuum description, and is inherently classical.
A discretized version of the skyrmion number is introduced in Ref.~\cite{BERGL} as the total solid angle spanned by the spins, divided by the solid angle of a sphere, $4\pi$. 
% In Ref.~\cite{Lohani}, the expression given in Ref.~\cite{LohaniSolidAngle} for the solid angle is used, and the spin products are replaced by correlation functions to obtain a quantum mechanical analog of the topological invariant \cite{Sotnikov, Lohani}.
In Ref.~\cite{Lohani}, the spin products are replaced by correlation functions to obtain a quantum mechanical analog of the topological invariant \cite{Sotnikov, Lohani}.
The expression is here adapted to general $S$ and we use the $\operatorname{atan2}$ function to get the correct value of the solid angle for all possible combinations of three spins \cite{LohaniSolidAngle}. The quantum skyrmion OP is
\begin{equation}
\label{eq:QBL}
    Q_{\text{BL}} = \frac{1}{2\pi} \sum_{\langle ijk \rangle} \operatorname{atan2}\big(\langle N_{ijk} \rangle , \langle D_{ijk} \rangle \big),
\end{equation}
where $\langle N_{ijk} \rangle = \langle \boldsymbol{S}_i \cdot (\boldsymbol{S}_j \cross \boldsymbol{S}_k) \rangle/S^3$ and $\langle D_{ijk} \rangle =  1+(\langle \boldsymbol{S}_i \cdot \boldsymbol{S}_j  \rangle + \langle \boldsymbol{S}_i \cdot \boldsymbol{S}_k \rangle + \langle \boldsymbol{S}_j \cdot \boldsymbol{S}_k  \rangle )/S^2$.
The sum is taken over all nonequivalent triangles of nearest neighbors, oriented counterclockwise \cite{Lohani}. When treating a discrete system of quantum spins, $Q_{\text{BL}}$ is analogous to the topological invariant of the corresponding classical GS. However, since it is no longer topological in a rigorous sense \cite{Sotnikov}, we name it the quantum skyrmion OP.

For the two SkXs in Fig.~\ref{fig:GS_spectrum}(a,c) the classical value $Q_{\text{BL},0}$ is an integer per magnetic unit cell, $Q_{\text{BL},0}/N' = 1$. The value confirms the presence of one skyrmion in the magnetic unit cell. For the degenerate states with all spins flipped compared to SkX1 and SkX2, $Q_{\text{BL},0}/N' = -1$. Since its sign does not distinguish skyrmions and antiskyrmions, it makes most sense to discuss the absolute value of $Q_{\text{BL}}$. We choose to work exclusively with the SkXs that have $Q_{\text{BL},0}/N' = 1$ in this paper. To avoid any confusion in the following we will not put absolute values on the quantum skyrmion OP.

The aim of this paper is to go beyond the classical, integer-valued $Q_{\text{BL},0}$ for the skyrmion OP.
%In Ref.~\cite{Sotnikov}, they used exact diagonalization of a 19-site cluster, defined the quantum scalar chirality as an approximation to $Q_{\text{BL}}$, and considered its utility as a quantum skyrmion OP. We study periodic lattices of skyrmions and use the HP approach to study quantum fluctuations around the classical GS. Here, the quantum scalar chirality would be $Q = (1/8\pi)\sum_{\langle ijk \rangle} \langle N_{ijk} \rangle$. 
In Ref.~\cite{Sotnikov}, they defined the quantum scalar chirality as an approximation to $Q_{\text{BL}}$, and considered its utility as a quantum skyrmion OP. Within the HP approach, the quantum scalar chirality would be $Q = (1/8\pi)\sum_{\langle ijk \rangle} \langle N_{ijk} \rangle$. 
The phases SkX1 and SkX2 contain triangles where $\langle N_{ijk} \rangle >0$ and $\langle D_{ijk} \rangle < 0$, and therefore the quantum scalar chirality is a poor approximation to $Q_{\text{BL}}$ in these SkXs. 
%In the HP approach we can calculate the quantum fluctuations in $Q_{\text{BL}}$ directly, and use that as the quantum skyrmion OP. The goal is to check whether $Q_{\text{BL}}$ is a good OP for quantum skyrmions. 
In the HP approach we can calculate the quantum fluctuations in $Q_{\text{BL}}$ directly, and then check whether $Q_{\text{BL}}$ is a good OP for quantum skyrmions. 

\begin{figure}
    \centering
    \includegraphics[width=0.9\linewidth]{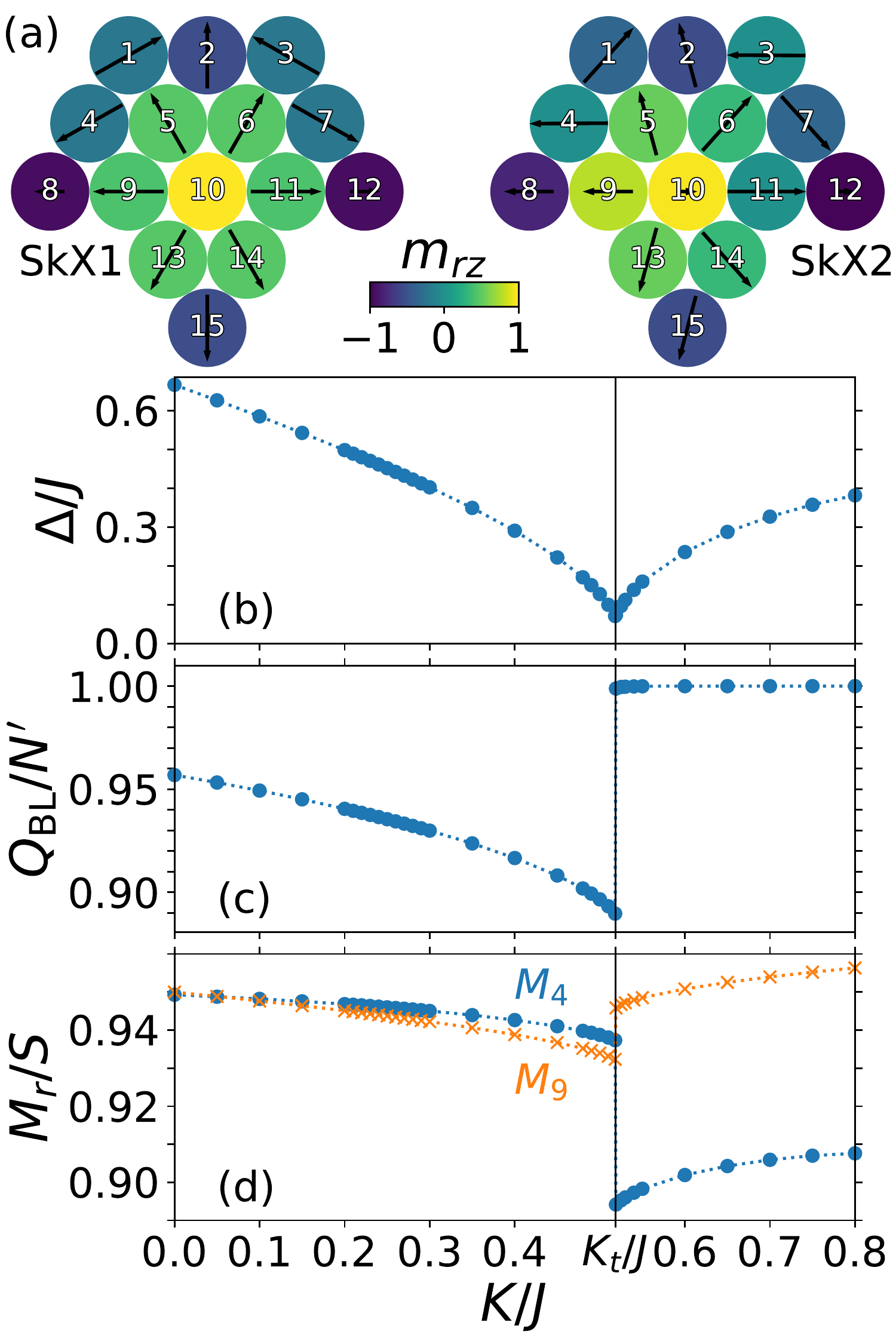}
    \caption{(a) The magnetic unit cells of SkX1 and SkX2 which are separated by a quantum phase transition (QPT) at $K = K_t$. The numbering of the sublattices is included, which was chosen so that the spin with the largest absolute value of $m_{rz}$ is sublattice $r = 10$. (b) The magnon gap $\Delta = \min_{\boldsymbol{k}} E_{\boldsymbol{k}}$. The dependence of the gap on the easy-axis anisotropy changes at the QPT. (c) Quantum skyrmion order parameter (OP), $Q_{\text{BL}}$, given per magnetic unit cell. The classical value is $1$ in both phases. Notice that the degree of quantum fluctuations increases when the magnon gap decreases. The quantum fluctuations are stronger in SkX1 compared to SkX2, and hence $Q_{\text{BL}}$ is a good OP to characterize the QPT. (d) $M_4$ and $M_9$ are shown, which are the sublattice magnetizations with the strongest fluctuations in SkX2 and SkX1, respectively. The parameters are $D/J = 2.16$, $U/J = 0.35$, and $S=1$. Calculated values are shown with markers, while dotted lines are added for visualization.}
    \label{fig:fluct}
\end{figure}

After inserting the HP transformation to second order in magnon operators, we write $N_{ijk} = N_{ijk, 0} + N_{ijk, 1} + N_{ijk, 2}$ and $D_{ijk} = D_{ijk, 0} + D_{ijk, 1} + D_{ijk, 2}$. The terms to zeroth order in magnon operators are the classical values. The expectation values of the linear terms vanish in the diagonalized frame, and so $\langle N_{ijk, 1} \rangle = \langle D_{ijk, 1} \rangle = 0$. The quadratic parts are
\begingroup
\allowdisplaybreaks
\begin{align}
    N_{ijk, 2} =& \frac{1}{2S} \Big(-2\hat{e}_3^{i} \cdot(\hat{e}_3^{j} \cross \hat{e}_3^{k})(a_i^\dagger a_i + a_j^\dagger a_j + a_k^\dagger a_k) \nonumber \\
    &+\big[\hat{e}_-^{i} \cdot(\hat{e}_-^{j} \cross \hat{e}_3^{k})a_i a_j +\hat{e}_-^{i} \cdot(\hat{e}_+^{j} \cross \hat{e}_3^{k})a_i a_j^\dagger  \nonumber \\
    &+\hat{e}_-^{i} \cdot(\hat{e}_3^{j} \cross \hat{e}_-^{k})a_i a_k +\hat{e}_-^{i} \cdot(\hat{e}_3^{j} \cross \hat{e}_+^{k})a_i a_k^\dagger \nonumber \\
    &+\hat{e}_3^{i} \cdot(\hat{e}_-^{j} \cross \hat{e}_-^{k})a_j a_k  +\hat{e}_3^{i} \cdot(\hat{e}_-^{j} \cross \hat{e}_+^{k})a_j a_k^\dagger \big] \nonumber \\
    &+ \text{H.c.} \Big), 
\end{align}
where H.c. denotes the Hermitian conjugate of the preceding term, and
\begin{align}
    D_{ijk, 2} =& \frac{1}{2S} \Big( -2 \hat{e}_3^{i} \cdot(\hat{e}_3^{j} + \hat{e}_3^{k}) a_i^\dagger a_i -2 \hat{e}_3^{j} \cdot(\hat{e}_3^{i} + \hat{e}_3^{k})a_j^\dagger a_j \nonumber \\&-2 \hat{e}_3^{k} \cdot(\hat{e}_3^{i} + \hat{e}_3^{j})a_k^\dagger a_k +\big[ \hat{e}_-^{i} \cdot \hat{e}_-^{j} a_i a_j \nonumber \\
    &+ \hat{e}_-^{i} \cdot \hat{e}_+^{j}a_i a_j^\dagger + \hat{e}_-^{i} \cdot \hat{e}_-^{k}a_i a_k + \hat{e}_-^{i} \cdot \hat{e}_+^{k} a_i a_k^\dagger \nonumber \\
    &+ \hat{e}_-^{j} \cdot \hat{e}_-^{k} a_j a_k + \hat{e}_-^{j} \cdot \hat{e}_+^{k} a_j a_k^\dagger \big]+ \text{H.c.} \Big). 
\end{align}
\endgroup
It is assumed that $i \in r, j \in s,$ and $k \in t$ to introduce the FT in these expressions. Simultaneously, the sum over sites $ijk$ is replaced by a sum over sublattices $rst$, including periodic boundary conditions (PBCs) \cite{PBChexagon}, times the number of magnetic unit cells. To calculate the expectation values, the transformation matrix in Eq.~\eqref{eq:ColpaT} is used to transform to the diagonal basis. Focusing on quantum fluctuations, the temperature is set to zero. Then, the only contribution comes from the commutator in the terms $\langle b_{\boldsymbol{k},n}b_{\boldsymbol{k},n}^\dagger \rangle = 1 + \langle b_{\boldsymbol{k},n}^\dagger b_{\boldsymbol{k},n} \rangle$, where $\langle b_{\boldsymbol{k},n}^\dagger b_{\boldsymbol{k},n} \rangle = 0$ at zero temperature. All other operator combinations give zero expectation value in the diagonalized basis. An example of a FT is $a_i a_j = \frac{1}{N'}\sum_{\boldsymbol{k}}\sum_{\boldsymbol{k}'} e^{i\boldsymbol{k} \cdot \boldsymbol{r}_i} a_{\boldsymbol{k}}^{(r)} e^{i\boldsymbol{k}' \cdot \boldsymbol{r}_j}a_{\boldsymbol{k}'}^{(s)}$. Only $\boldsymbol{k}' = -\boldsymbol{k}$ contributes to the expectation value, and so we get a factor $e^{i\boldsymbol{k}\cdot (\boldsymbol{r}_i-\boldsymbol{r}_j)} = e^{-i\boldsymbol{k}\cdot \boldsymbol{\delta}_{(r,s)}}$. For more details of how the expectation values are calculated, see Appendix \ref{app:colpa}.

We calculate the expectation values $\langle N_{rst} \rangle$ and $\langle D_{rst} \rangle$ for the 30 nonequivalent triangles involving nearest-neighbor sites at sublattices $rst$, and then insert these in $Q_{\text{BL}}/N' = (1/2\pi) \sum_{\langle rst \rangle} \operatorname{atan2}(\langle N_{rst} \rangle , \langle D_{rst} \rangle )$. The results for $Q_{\text{BL}}/N'$ are shown in Fig.~\ref{fig:fluct}(c) as a function of the easy-axis anisotropy $K$. We see that $Q_{\text{BL}}/N'$ remains clearly nonzero, and so is a good quantum skyrmion OP. It is evident that the quantum fluctuations do not destroy the skyrmion nature of the GS. The jump in $Q_{\text{BL}}/N'$ at the transition point $K_t$ shows that it is also a good OP for the QPT between two quantum skyrmion phases, SkX1 and SkX2. This is in contrast to the classical version which is the same in both phases. The jump in the OP indicates that the transition between SkX1 and SkX2 is a discontinuous QPT and not a quantum critical phenomenon. 

In SkX2, $Q_{\text{BL}}/N'$ approaches 1 from below with increasing $K$, indicating that the quantum fluctuations in SkX2 barely affect its skyrmion nature. SkX1 contains four triangles in the magnetic unit cell where $\langle D_{rst}\rangle < 0$, while SkX2 contains only two such triangles. In reference to the sublattice numbering in Fig.~\ref{fig:fluct}(a) an example of a triangle with  $\langle D_{rst}\rangle < 0$ that applies to both phases is $r=6, s = 7, t = 3$. In SkX1 at $K/J = 0.50$ we find that $N_{rst, 0} \approx 0.36$, $\langle N_{rst, 2}\rangle \approx 0.02$,  $D_{rst, 0} \approx -0.11$, and $\langle D_{rst, 2} \rangle \approx 0.06$ in these triangles. Hence, the quantum fluctuations have a significant effect on their contribution to the total solid angle. Meanwhile, in SkX2, there are two triangles where $0 <  N_{rst,0} \ll 1$, $0 < \langle N_{rst,2} \rangle \ll N_{rst,0}$, $D_{rst, 0} \approx -0.47$, and $\langle D_{rst, 2} \rangle \approx 0.07$ for $K/J = 0.53$. Therefore, the quantum fluctuations have a negligible effect on their contribution to the total solid angle. While there are 30 triangles contributing to $Q_{\text{BL}}/N'$, we suggest this difference in the subset of triangles with $\langle D_{rst}\rangle < 0$ explains the much smaller effect of quantum fluctuations on the quantum skyrmion OP in SkX2 compared to SkX1.

The quantum fluctuations in the OP, i.e., the deviation of $Q_{\text{BL}}/N'$ from the classical value $Q_{\text{BL}, 0}/N' = 1$, increases with decreasing magnon gap, see Fig.~\ref{fig:fluct}(b,c). This is understood from the fact that quantum fluctuations become less energetically costly with lower magnon gap. In Fig.~\ref{fig:fluct} we employ a higher density of $K/J$ values between $0.2$ and $0.3$ because there are interesting topological magnon features in this region \cite{upcoming}. We find that these features do not affect the quantum fluctuations. A higher density is also used around $K_t$ where the gap and the fluctuations change more rapidly as functions of $K$. It is emphasized that new GSs are found at each value of $K$.

The classical version, $Q_{\text{BL},0}$, is zero for all coplanar states, except for a set of exceptional states where three neighboring spins give $N_{ijk,0} = 0$ and $D_{ijk,0} < 0$. As pointed out in Ref.~\cite{BERGL}, $Q_{\text{BL},0}$ is not defined in such a case because the sign of the spanned solid angle of $2\pi$ is ambiguous. If one considers a system where such exceptional states are possible, one would need to consider whether such coplanar states are special cases of skyrmions, or, adjust the definition of $Q_{\text{BL},0}$ so that all possible coplanar states give $Q_{\text{BL},0} = 0$. Our results indicate that the fluctuations $\langle N_{ijk, 2} \rangle, \langle D_{ijk,2} \rangle $ always have smaller absolute values than the classical values $N_{ijk, 0}, D_{ijk, 0}$, even when $|N_{ijk, 0}| \ll 1$. Hence, a coplanar state which has $Q_{\text{BL},0} = 0$, will likely give $Q_{\text{BL}} = 0$ also when including quantum fluctuations. Therefore, $Q_{\text{BL}}$ should in general be a good OP for the existence of quantum skyrmions.

Probing $Q_{\text{BL}}$ in an experiment could involve measuring the spin correlation functions in Eq.~\eqref{eq:QBL}. The spin-spin correlation functions in $\langle D_{ijk} \rangle$ can be measured by neutron scattering \cite{MeasureSpinCorrNeutron_1968, MeasureSpinCorrNeutron} or magnetostriction experiments \cite{MeasureSpinCorrelation}. Information about the FT of the three-spin correlation function $\langle N_{ijk} \rangle$, also referred to as the scalar spin chirality, can be extracted for very small momenta using Raman scattering \cite{3spinRamanTheory, 3spinRamanExp}. Proposals to access information about the FT of the scalar spin chirality at finite momenta include neutron scattering \cite{3spinNeutron} and resonant inelastic x-ray scattering \cite{3spinXray}.

%-----------------------magnetization, energy correction---------------------
\section{Magnetization} \label{sec:fluct}
We define the magnetization of sublattice $r$ as
\begin{equation}
    M_r = \frac{1}{N'}  \sum_{i\in r} \left\langle S_{i3} \right\rangle = S - \frac{1}{N'}\sum_{\boldsymbol{k}} \langle a_{\boldsymbol{k}}^{(r)\dagger} a_{\boldsymbol{k}}^{(r)} \rangle.
\end{equation}
In SkX1 we find the strongest quantum fluctuations for the two sublattices to the left and to the right of the central up spin. In our chosen numbering these are sublattices $9$ and $11$, see Fig.~\ref{fig:fluct}(a). In SkX2 sublattices $3$ and $4$ show the strongest quantum fluctuations. We show $M_4$ and $M_9$ as functions of $K$ for both phases in Fig.~\ref{fig:fluct}(d). We see that the corrections to the classical value $M_r = S$ are less than $11\%$, supporting the validity of the HP approach. Again, we see that the fluctuations increase with decreasing gap. The spins at sublattice 10 show the weakest quantum fluctuations in both phases, which remain less than $1\%$ for all considered $K$.

%--------------------------------CONCLUSION----------------------------------------------
\section{Conclusion} \label{sec:Con}
We studied a weakly exchange-coupled magnetic monolayer where the Dzyaloshinskii-Moriya interaction and the four-spin interaction conspire to create a dense skyrmion crystal. A quantum phase transition between two distinct skyrmion crystals, driven by a tunable easy-axis anisotropy, was found. The magnon spectrum was derived and discussed for both quantum skyrmion phases. The discrete and quantum natures of small skyrmions are not negligible, and hence the continuum and classical topological invariant is rendered ill-defined. We proceeded to study the quantum fluctuations in the discretized order parameter of quantum skyrmions. The order parameter proved to be a good indicator of the presence of quantum skyrmions. Furthermore, it managed to capture the quantum phase transition between distinct skyrmion phases, signaled by a jump in the order parameter at a particular value of the easy-axis anisotropy. It was found that quantum fluctuations do not adversely affect the skyrmion nature of quantum skyrmions compared to their classical ground states.

%--------------------------------ACKNOWLEDGEMENTS-------------------------------------------
\section*{Acknowledgments}
We acknowledge funding from the Research Council of Norway through its Centres of Excellence funding scheme, Project No.~262633, ``QuSpin," and through Project No.~323766, ``Equilibrium and out-of-equilibrium quantum phenomena in superconducting hybrids with antiferromagnets and topological insulators."

%We acknowledge funding from the Research Council of Norway through Project No.~323766, ``Equilibrium and out-of-equilibrium quantum phenomena in superconducting hybrids with antiferromagnets and topological insulators," and the Research Council of Norway  through its Centres of Excellence funding scheme, Project No.~262633, ``QuSpin." 

%---------------------------APPENDICES---------------------------------------
\appendix
\section{Ground states} \label{app:GS}
%Obtaining the ground state, tuning parameters, periodicity of SkX, GS similar to the SkX measured in Heinze experiment. Details in appendix.
\subsection{Periodicity}
To obtain the classical GS we replace the spin operators in the Hamiltonian in Eq.~\eqref{eq:H} by $\boldsymbol{S}_i = S \boldsymbol{m}_i$, where $S$ is the spin quantum number, and $\boldsymbol{m}_i$ is a unit vector determining the direction of the spin at lattice site $i$. The GS is then the set $\{\boldsymbol{m}_i\}$ which minimizes $H(\{\boldsymbol{m}_i\})$. A challenge is that the periodicity of potential SkXs is a priori unknown, and so any calculation involving a finite number of lattice sites with PBCs will suffer from finite-size effects. We therefore first determine the parameters which give a SkX with the desired periodicity, before finding the best state with that periodicity.

The approach starts from the SkX construction detailed in the supplementary material of Ref.~\cite{HeinzeSkX}. Among the SkX states covered there, the one which gives the lowest energy in our model is used as a trial state and named SkXt. SkXt is constructed using $\boldsymbol{Q}_M =  2\pi \hat{x}/\lambda_x, \boldsymbol{Q}_K =  2\pi \hat{y}/ (\lambda_y\sqrt{3}/2), \Tilde{\phi}_i = \boldsymbol{Q}_M \cdot \boldsymbol{r}_i, \Tilde{\theta}_i = \boldsymbol{Q}_K \cdot \boldsymbol{r}_i,$ and $\boldsymbol{m}_i = (\sin\Tilde{\phi}_i \cos\Tilde{\theta}_i/|\cos\Tilde{\theta}_i|, \cos\Tilde{\phi}_i \sin\Tilde{\theta}_i, \cos\Tilde{\phi}_i \cos\Tilde{\theta}_i)$. Here $\lambda_x$ is the periodicity in the $x$ direction in units of the lattice constant, while $\lambda_y$ is the periodicity in the $y$ direction in units of $\sqrt{3}/2$ lattice constants. SkXt arranges skyrmions in a centered rectangular lattice. The periodicities $\lambda_x$ and $\lambda_y$ used to specify the GS refer to the conventional unit cell of the centered rectangular lattice, which contains two skyrmions here. It is possible to consider any rational values of $\lambda_x$ and $\lambda_y$ and still make finite-sized lattices with correct PBCs. 

\begin{figure}
    \centering
    \includegraphics[width=\linewidth]{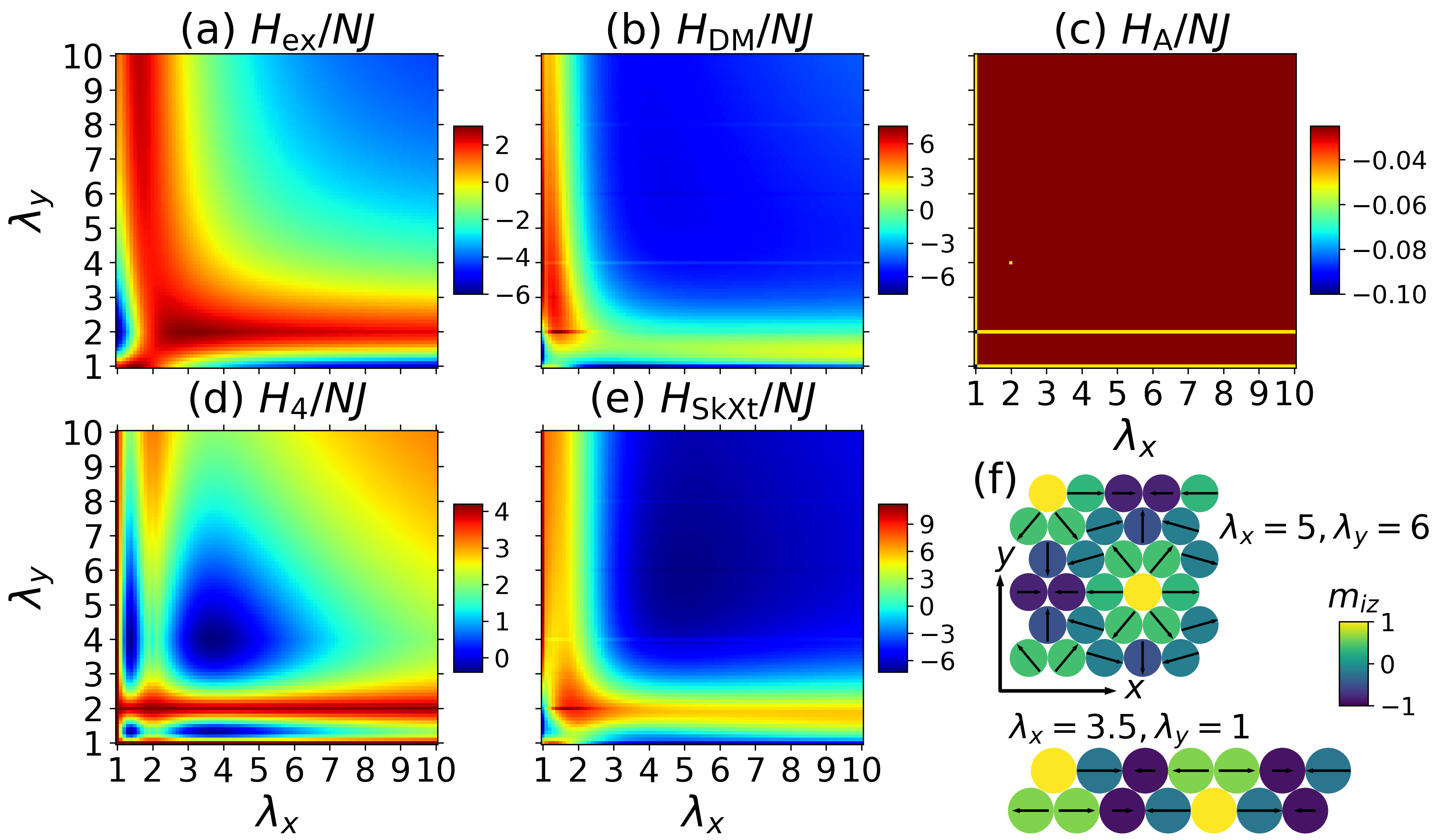}
    \caption{(a-d) The individual contributions to the classical Hamiltonian of the trial skyrmion crystal state SkXt as a function of the periodicities in the $x$ and $y$ directions. The energies are given per lattice site, and were calculated with periodicity-dependent lattice sizes in order to implement correct periodic boundary conditions. (e) The total energy of SkXt, which is minimal for $\lambda_x = 5, \lambda_y = 6$. (f) SkXt for $\lambda_x = 5, \lambda_y = 6$ as well as $\lambda_x = 3.5, \lambda_y = 1$ which is a helical state that is strongly favored by $H_{\text{DM}}$ and strongly disfavored by $H_4$. The parameters are $D/J = 2.16$, $U/J = 0.35$, $K/J = 0.10$, and $S=1$.}
    \label{fig:periodicity}
\end{figure}

Fig.~\ref{fig:periodicity} shows the four individual contributions, as well as the total classical Hamiltonian for SkXt. The exchange and easy-axis anisotropy terms, see Fig.~\ref{fig:periodicity}(a,c), prefer a ferromagnetic GS, which is achieved with $\lambda_x = 1, \lambda_y = 2$. 
%Also notice that the exchange interaction favors large periodicities, since then neighboring spins are more aligned. 
Special values $\lambda_x = 1$ and general $\lambda_y$, as well as $\lambda_y = 1$ or $\lambda_y = 2$ and general $\lambda_x$, give helical states. Some of these are favored by the DMI contribution, see Fig.~\ref{fig:periodicity}(b), while all are disfavored by the four-spin interaction, see Fig.~\ref{fig:periodicity}(d). In addition, $\lambda_x = 2, \lambda_y = 4$ gives a coplanar state, which, along with the helical states (also coplanar), yields a lower $H_{\text{A}}$ compared to the noncoplanar skyrmion-like states, see Fig.~\ref{fig:periodicity}(c). Among the possible SkXt states, the DMI term prefers a helical state with $\lambda_x \approx 3.4, \lambda_y = 1$. Fig.~\ref{fig:periodicity}(f) shows the similar helical state with $\lambda_x = 3.5, \lambda_y = 1$. 

When focusing on SkXt states, it is clear that the value of the easy-axis anisotropy has no effect on the periodicity unless it is the dominant energy in the system. However, the true GS will adjust the $z$ components of its spins to take advantage of an increased $K$. We kept $K$ in the region $0 \leq K/J \leq 0.8$ in this paper and ensured that it did not affect the periodicity of the GS. The total classical Hamiltonian for SkXt states, see Fig.~\ref{fig:periodicity}(e), is minimized for $\lambda_x = 5, \lambda_y = 6$ at $D/J = 2.16$ and $US^2/J = 0.35$, and that SkXt state is shown in Fig.~\ref{fig:periodicity}(f). To be specific, we tuned the parameters to ensure that $|\lambda_x-5|, |\lambda_y-6| < 0.001$ minimizes the total energy of SkXt. This also shows how $U/J$ must be tuned to get the same GS if one considers different $S$. We have set $S=1$ and $U/J = 0.35$ in this paper. 

\subsection{Obtaining the ground states}
We employ two methods of searching for the GSs, namely Monte Carlo simulated annealing \cite{simulatedannealing, SkMCOkubo, SkMCRosza, DuineMCHPSk, SkMCsiemens, HeinzeSkXMC, Rosales, DiazPRL, DiazPRR} and self-consistent iteration \cite{dosSantosPRB, RoldanMolina, RoldanMolina2016}. We obtained the lowest-energy states using the iterative approach, and therefore explain it in detail here. Knowing the periodicity, we can work with the 15-site magnetic unit cell with PBCs \cite{PBChexagon}, see Fig.~\ref{fig:fluct}(a) and compare to Fig.~\ref{fig:GS_spectrum}(a,c). The magnetization on each site is set to $\boldsymbol{m}_i = (\sin\theta_i\cos\phi_i, \sin\theta_i\sin\phi_i, \cos\theta_i)$, where the inclination $\theta_i$ and the azimuth $\phi_i$ are the familiar angles in spherical coordinates. We start from both a random set of angles and the trial state SkXt. Let $n \in \mathbb{N}$ be the number of iterations completed. Then, for each spin the magnetic torques, $T_i^\theta = \partial_{\theta_i^n}{H}$ and $T_i^\phi = \partial_{\phi_i^n}{H}$, are calculated. These are used to set new angles $\theta_i^{n+1} = \theta_i^{n} -\alpha T_i^\theta$ and $\phi_i^{n+1} = \phi_i^{n} -\alpha T_i^\phi ,$ where $\alpha$ is the mixing parameter \cite{dosSantosPRB}. We used $\alpha J = 0.001$ and $0.0001$. The iterations are repeated until self-consistency is reached.

%\hyperref[fig:GS_spectrum]{\ref*{fig:GS_spectrum}(a)}
At $K/J = 0.518$ the end result is the SkX1 state shown in Fig.~\ref{fig:GS_spectrum}(a), while at  $K/J = 0.519$ the end result is the SkX2 state shown in Fig.~\ref{fig:GS_spectrum}(c). 
We emphasize that the SkXs we find as GSs have lower energy than the trial states SkXt used to determine the periodicity. The 15 sublattices in the GSs are equal, centered rectangular lattices with primitive vectors $\boldsymbol{a}_1 = (5/2, -3\sqrt{3}/2)$ and $\boldsymbol{a}_2 = (5/2, 3\sqrt{3}/2)$. Their 1BZ is a nonregular hexagon with vertices at $(\pm 52\pi/135, 0)$ and $(\pm 2\pi/135,\pm 2\pi/3\sqrt{3})$. We define the points $\boldsymbol{\Gamma} = (0,0)$, $\boldsymbol{X} = (52\pi/135, 0)$, $\boldsymbol{M} = (\pi/5, \pi/3\sqrt{3})$, $\boldsymbol{S} = (2\pi/135, 2\pi/3\sqrt{3})$, and $\boldsymbol{Y} = (0, -2\pi/3\sqrt{3})$ in the 1BZ, as shown in Fig.~\ref{fig:GS_spectrum}(f).

We performed several checks on the proposed GSs. Starting from random states on larger lattices and using Monte Carlo did not yield any states with lower energy. More convincingly, it can be shown that requiring the coefficients of linear terms in the Hamiltonian to be zero, is equivalent to requiring that the zero-order terms $H_0$ are in an extremum, $\partial_{\theta_i}{H_0} = \partial_{\phi_i}{H_0} = 0$. 
%In the GS, the coefficients of linear terms should be zero. 
By calculating the coefficients of the linear terms from the proposed GSs, we get zero within numerical accuracy. 

In the numerics, we set $\hbar = a = J = S = 1$ so that all quantities are dimensionless. In the numerical GSs obtained with 15 free spins in the magnetic unit cell, the spins show small, $\order{10^{-7}}$, deviations from the symmetries discussed in Sec.~\ref{sec:spectrum}. By requiring these apparent symmetries to be exact, we find alternative GSs with the same energy to $\order{10^{-15}}$. These small deviations from the apparent symmetries of the classical GSs are therefore viewed as numerical artifacts, which have minor effects on the results. The maximum absolute values of the coefficients of linear terms in the Hamiltonian are $\order{10^{-7}}$ due to these artifacts, while exactly symmetric GSs give $\order{10^{-10}}$ coefficients of linear terms. All numerical results presented in this paper are accurate down to $\order{10^{-7}}$ or better. Achieving accuracy down to $\order{10^{-10}}$ should be possible by requiring the numerical GSs to be exactly symmetric, but the corrections would not be visible in any plots presented here.

\subsection{Energy correction and net magnetization}

\begin{figure}
    \centering
    \includegraphics[width=0.9\linewidth]{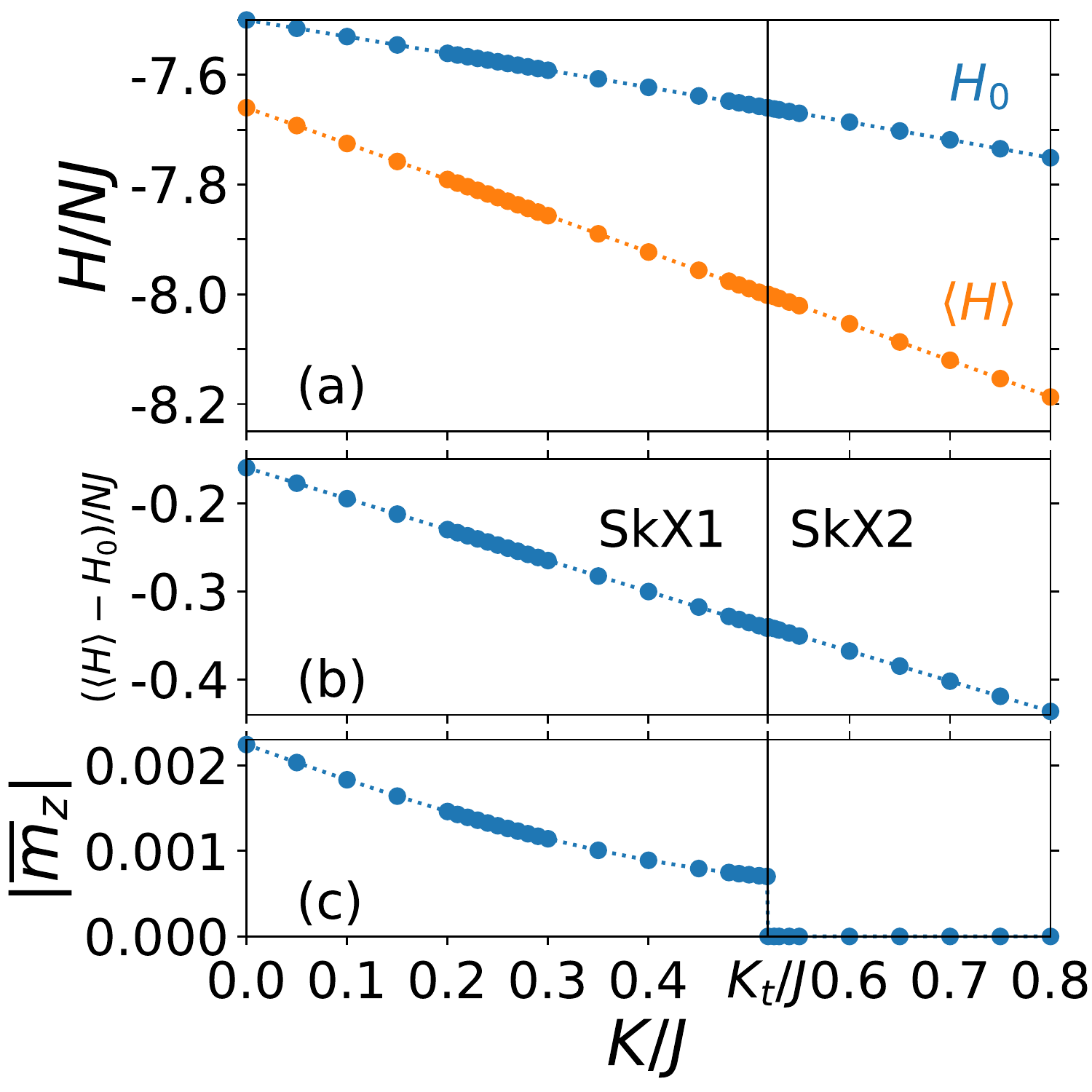}
    \caption{(a) The classical ground state (GS) energy $H_0$ and the expectation value of the Hamiltonian up to second order in magnon operators, $\langle H \rangle$, given per lattice site. Their difference is shown in (b). In (c) the net magnetization $\overline{m}_{z}$ of the GS is shown as a function of the easy-axis anisotropy $K$. Notice the jump at the QPT between SkX1 and SkX2. The parameters are the same as those in Fig.~\ref{fig:fluct}.}
    \label{fig:encorr}
\end{figure}

Bosonic commutators were used to write the quadratic terms in the Hamiltonian on the form in Eq.~\eqref{eq:H2}, and these yield quantum corrections to the GS energy. Also, the diagonalized $H_2$ in Eq.~\eqref{eq:H2diag} gives a nonzero contribution at zero temperature, and so the expectation value of the Hamiltonian is
\begin{align}
    \langle H \rangle = H_0 + \frac{N'}{2} \sum_r \pqty{\nu_r - \eta_r} + \frac{1}{2}\sum_{\boldsymbol{k}}\sum_{n = 1}^{15} E_{\boldsymbol{k}, n},
\end{align}
where $H_0$ is the classical GS energy. We plot $H_0$ and $\langle H \rangle$ in Fig.~\ref{fig:encorr}(a). Notice that quantum corrections lower the energy of the skyrmions, which means that quantum fluctuations stabilize skyrmion crystals. Hence, it is the quantum skyrmions, and not the classical GSs, which are the preferred states of the system. This is completely analogous to the HP treatment of antiferromagnets. The same has previously been found for skyrmions in a magnetic field \cite{RoldanMolina, Sotnikov}. 
%Along with their small size, this is our justification for referring to the skyrmions in SkX1 and SkX2 as quantum skyrmions.
The plot of $\langle H \rangle - H_0$ in Fig.~\ref{fig:encorr}(b) is much closer to being linear than the other plots of quantum fluctuations in Fig.~\ref{fig:fluct}. By inspecting the calculated values we find that the magnitude of the slope increases slightly for increasing $K$ in both phases. It appears that changes in $\langle H \rangle$ are dominated by the fact that the Hamiltonian depends linearly on $K$. Also, gradual changes in the GS to take advantage of higher $K$ are more important than the value of the magnon gap, since even in SkX2, where the gap is an increasing function of $K$, the magnitude of the slope of $\langle H \rangle - H_0$ increases with $K$.

%In the main text, we determined the critical easy-axis anisotropy $K_c$ of the QPT between SkX1 and SkX2 from the classical GS energy $H_0$. One could also include quantum fluctuations of the energy, and use $\langle H \rangle$ to determine the critical $K$. SkX1 remains a local minimum in the classical $H_0$ for $K>K_c$, and hence the linear terms in the Hamiltonian remain zero. Therefore, it is possible to use updated versions of SkX1 at $K>K_c$ and perform calculations of the quantum fluctuation around these states. We then find that also for $K>K_c$ SkX1 has lower $\langle H \rangle $ than SkX2. However, at $K = \overline{K}_c$, with $0.526 < \overline{K}_c/J < 0.527$, the gap in the magnon spectrum of SkX1 closes. The matrix $M_{\boldsymbol{k}}$ in Eq.~\eqref{eq:M} is no longer positive definite for SkX1, and so the method in Ref.~\cite{Tsallis} must be used to diagonalize the system \cite{COLPA}. This yields complex energy eigenvalues for SkX1 at $K > \overline{K}_c$ indicating that SkX1 becomes dynamically unstable \cite{PethickSmith}. Hence, if determined using $\langle H \rangle$ rather than $H_0$, SkX1 is the preferred GS up to $K = \overline{K}_c$, at which point a QPT to the dynamically stable SkX2 takes place. 

In the main text, we determined the transition point $K = K_t$ of the QPT between SkX1 and SkX2 from the classical GS energy $H_0$. One could also include quantum fluctuations of the energy, and use $\langle H \rangle$ to determine the transition point. SkX1 remains a local minimum in the classical $H_0$ for $K>K_t$, and hence the linear terms in the Hamiltonian remain zero. Therefore, it is possible to use updated versions of SkX1 at $K>K_t$ and perform calculations of the quantum fluctuations around these states. We then find that also for $K>K_t$ SkX1 has lower $\langle H \rangle $ than SkX2. However, at $K = \Tilde{K}_t$, with $0.526 < \Tilde{K}_t/J < 0.527$, the gap in the magnon spectrum of SkX1 closes. The matrix $M_{\boldsymbol{k}}$ in Eq.~\eqref{eq:M} is no longer positive definite for SkX1, and so the method in Ref.~\cite{Tsallis} must be used to diagonalize the system \cite{COLPA}. This yields complex energy eigenvalues for SkX1 at $K > \Tilde{K}_t$ indicating that SkX1 becomes dynamically unstable \cite{PethickSmith}. Hence, if determined using $\langle H \rangle$ rather than $H_0$, SkX1 is the preferred state up to $K = \Tilde{K}_t$, at which point a QPT to the dynamically stable SkX2 takes place.

The SkX1 GSs have a small net magnetization, $\overline{m}_{z} = \sum_i m_{iz}/N$, as opposed to the SkX2 states and the trial SkXt state which have zero net magnetization. The net magnetization is plotted in Fig.~\ref{fig:encorr}(c), and it is negative for the SkX1 states with spin up at the center. The Hamiltonian in Eq.~\eqref{eq:H} does not have Ising symmetry but it would still be natural to assume that the net magnetization of a SkX GS is zero, since there is no external magnetic field. We believe the small net magnetization of SkX1 is simply due to the fact that such a configuration allows a lower total energy from the four competing contributions to the Hamiltonian when placing a dense SkX on the triangular lattice.

%-----------------------COLPA-------------------------------------------
\section{Diagonalization method} \label{app:colpa}
The matrix $M_{\boldsymbol{k}}$ in Eq.~\eqref{eq:M} is Hermitian, $M_{\boldsymbol{k}}^\dagger = M_{\boldsymbol{k}}$, while its block elements obey $\eta_{\boldsymbol{k}}^\dagger = \eta_{\boldsymbol{k}}$ and $\nu_{\boldsymbol{k}}^T = \nu_{-\boldsymbol{k}}$. $M_{\boldsymbol{k}}$ is also positive definite in this bosonic system. Therefore, we employ the matrix generalization of the Bogoliubov transformation introduced in Ref.~\cite{COLPA} to diagonalize the Hamiltonian. Then, we are guaranteed that the excitation spectrum is real \cite{COLPA}, and hence that the system is stable \cite{PethickSmith}.
In Ref.~\cite{PWSWBEC}, we used the method described in Refs.~\cite{Tsallis, Xiao} because the matrix $M_{\boldsymbol{k}}$ was not positive definite in all phases. We have confirmed that all results in this paper are the same when using either diagonalization method. 

The diagonalization \cite{COLPA},
\begin{equation*}
    \boldsymbol{a}_{\boldsymbol{k}}^\dagger M_{\boldsymbol{k}} \boldsymbol{a}_{\boldsymbol{k}} = (\boldsymbol{a}_{\boldsymbol{k}}^\dagger T_{\boldsymbol{k}}^\dagger)[(T_{\boldsymbol{k}}^{-1})^\dagger M_{\boldsymbol{k}} T_{\boldsymbol{k}}^{-1}] (T_{\boldsymbol{k}} \boldsymbol{a}_{\boldsymbol{k}}) = \boldsymbol{b}_{\boldsymbol{k}}^\dagger D_{\boldsymbol{k}} \boldsymbol{b}_{\boldsymbol{k}},
\end{equation*}
where $D_{\boldsymbol{k}}$ is diagonal, preserves the bosonic commutator relations for the diagonalized operators $ \boldsymbol{b}_{\boldsymbol{k}} = (b_{\boldsymbol{k},1}, \dots, b_{\boldsymbol{k},m}, b_{-\boldsymbol{k},1}^\dagger \dots b_{-\boldsymbol{k},m}^\dagger)^T$. 
%This method is permitted when $M_{\boldsymbol{k}}$ is positive definite, which it is in the system studied in this paper. Then, we are guaranteed that the excitation spectrum will be real \cite{COLPA}, and hence that the system is stable \cite{PethickSmith}. Another advantage of this method over the one in Ref.~\cite{Tsallis} is that it reduces the size of the matrices by half. 
For a system with $m$ degrees of freedom, the matrices are of size $2m \cross 2m$. In our system, $m = 15$, i.e., the number of sublattices in the GS. 

Unlike fermionic systems, the excitation spectrum is not simply the eigenvalues of $M_{\boldsymbol{k}}$, and the transformation matrix $T_{\boldsymbol{k}}$ is paraunitary, $T_{\boldsymbol{k}}^{-1} = \mathcal{J} T_{\boldsymbol{k}}^\dagger \mathcal{J}$ \cite{COLPA}, rather than unitary. Here, $\mathcal{J}$ is a diagonal matrix whose first $m$ diagonal elements are $1$, and final $m$ diagonal elements are $-1$. Since $M_{\boldsymbol{k}}$ is positive definite, its Cholesky decomposition, $M_{\boldsymbol{k}} = K_{\boldsymbol{k}}^\dagger K_{\boldsymbol{k}}$, exists, where $K_{\boldsymbol{k}}$ is upper triangular. The algorithm presented in Ref.~\cite{COLPA} then proceeds to find orthonormal eigenvectors $\boldsymbol{w}_{\boldsymbol{k},1}, ..., \boldsymbol{w}_{\boldsymbol{k},2m}$ and eigenvalues $E_{\boldsymbol{k},1}, ..., E_{\boldsymbol{k},2m}$ of the Hermitian matrix $K_{\boldsymbol{k}}\mathcal{J}K_{\boldsymbol{k}}^\dagger$. It can be shown that $E_{\boldsymbol{k}, n} = -E_{-\boldsymbol{k},n+m}$ for $n \leq m$, where $E_{\boldsymbol{k}, n} > 0$ and $E_{-\boldsymbol{k},n+m} < 0$. The $m$ positive eigenvalues are the excitation spectrum of the system \cite{COLPA}. To complete the diagonalization, the unitary matrix $W_{\boldsymbol{k}} = [\boldsymbol{w}_{\boldsymbol{k},1}| ...|\boldsymbol{w}_{\boldsymbol{k}, 2m}]$ is constructed from the orthonormal eigenvectors and the diagonal matrix $D_{\boldsymbol{k}} = \operatorname{diag}(E_{\boldsymbol{k},1}, ..., E_{\boldsymbol{k},m}, -E_{\boldsymbol{k},m+1}, \dots, -E_{\boldsymbol{k},2m})$ is constructed from the eigenvalues $E_{\boldsymbol{k},n}$. The inverse of the transformation matrix, $T_{\boldsymbol{k}}^{-1}$, is then calculated row by row from $K_{\boldsymbol{k}}T_{\boldsymbol{k}}^{-1} = W_{\boldsymbol{k}}D_{\boldsymbol{k}}^{1/2}$ starting at the last row. 

The transformation matrix can be written
\begin{equation}
\label{eq:ColpaT}
    T_{\boldsymbol{k}} = \begin{pmatrix} U_{\boldsymbol{k}} & V_{-\boldsymbol{k}}^* \\ V_{\boldsymbol{k}} & U_{-\boldsymbol{k}}^*    \end{pmatrix}.
\end{equation}
Then, using $\boldsymbol{a}_{\boldsymbol{k}} = T_{\boldsymbol{k}}^{-1} \boldsymbol{b}_{\boldsymbol{k}}$ the transformation to the diagonal basis is given by
\begingroup
\allowdisplaybreaks
\begin{align}
    a_{\boldsymbol{k}}^{(r)} =& \sum_{n = 1}^{m} \pqty{U_{\boldsymbol{k},r,n}^\dagger b_{\boldsymbol{k},n} -V_{\boldsymbol{k},r,n}^\dagger b_{-\boldsymbol{k},n}^\dagger }, \\
    a_{-\boldsymbol{k}}^{(r)\dagger} =& \sum_{n = 1}^{m} \pqty{-V_{-\boldsymbol{k},r,n}^T b_{\boldsymbol{k},n} +U_{-\boldsymbol{k},r,n}^T b_{-\boldsymbol{k},n}^\dagger }.
\end{align}
\endgroup
From this transformation, we find that at zero temperature
\begingroup
\allowdisplaybreaks
\begin{align}
    \langle a_{\boldsymbol{k}}^{(r)\dagger} a_{\boldsymbol{k}}^{(r)} \rangle =& \sum_{n = 1}^{m} V_{\boldsymbol{k},r,n}^T V_{\boldsymbol{k},r,n}^\dagger,\\
    \langle a_{\boldsymbol{k}}^{(r)} a_{-\boldsymbol{k}}^{(s)} \rangle =& \sum_{n = 1}^{m} U_{\boldsymbol{k},r,n}^\dagger (-V_{-\boldsymbol{k},s,n}^\dagger) , \\
    \langle a_{\boldsymbol{k}}^{(r)} a_{\boldsymbol{k}}^{(s)\dagger} \rangle =& \sum_{n = 1}^{m} U_{\boldsymbol{k},r,n}^\dagger U_{\boldsymbol{k},s,n}^T  ,
\end{align}
\endgroup
$\langle a_{\boldsymbol{k}}^{(r)\dagger} a_{-\boldsymbol{k}}^{(r)} \rangle = 0$, $ \langle a_{\boldsymbol{k}}^{(r)} a_{\boldsymbol{k}}^{(s)} \rangle = 0$, and $ \langle a_{\boldsymbol{k}}^{(r)} a_{-\boldsymbol{k}}^{(s)\dagger} \rangle = 0$. These are used extensively in the paper to calculate quantum fluctuations.

%--------------------Nonzero temperature------------------------------
\section{Thermal effects on order parameter} \label{app:Thermal}

To include thermal fluctuations when calculating expectation values of operator combinations, one also takes into account that
\begin{equation}
    \langle b_{\boldsymbol{k},n}^\dagger b_{\boldsymbol{k},n} \rangle = B_{\text{E}}(E_{\boldsymbol{k},n}) = \frac{1}{e^{ E_{\boldsymbol{k},n}/T}-1}    
\end{equation}
according to Bose-Einstein statistics. Here, $T$ is the temperature, and we have set Boltzmann's constant $k_{\text{B}} = 1$. As an example,
\begin{align*}
    \langle a_i a_j \rangle =& \frac{1}{N'}\sum_{\boldsymbol{k}}\sum_{n = 1}^{m} \Big( e^{i\boldsymbol{k}\cdot \boldsymbol{\delta}_{(r,s)}}(-V_{-\boldsymbol{k},r,n}^\dagger)U_{\boldsymbol{k}, s, n}^\dagger B_{\text{E}}(E_{\boldsymbol{k},n})    \nonumber \\
    &+ e^{-i\boldsymbol{k}\cdot \boldsymbol{\delta}_{(r,s)}} U_{\boldsymbol{k},r,n}^\dagger (-V_{-\boldsymbol{k},s,n}^\dagger) \bqty{1 + B_{\text{E}}(E_{\boldsymbol{k},n})} \Big) .
\end{align*}

Table \ref{tab:Thermal}(a) shows the results for the quantum skyrmion OP in SkX1 at $K/J = 0.10$. There, the magnon gap is $\Delta/J \approx 0.59$. At least when $T \leq \Delta$ the thermal effects do not adversely affect the utility of $Q_{\text{BL}}$ in indicating the presence of quantum skyrmions. The conclusion is the same when considering a lower magnon gap in SkX1 in Table \ref{tab:Thermal}(b). In fact, it seems thermal fluctuations have a lesser effect when viewed in terms of $T/\Delta$ here. This is due to the fact that the valley around the global minimum of $E_{\boldsymbol{k},15}$ is sharper at $K/J = 0.50$ than at $K/J = 0.10$. Like the quantum fluctuations, SkX2 is also less affected by thermal fluctuations compared to SkX1, see Table \ref{tab:Thermal}(c). We suggest the reason is the same as that given when discussing the quantum fluctuations in Sec.~\ref{sec:OP}. If the weak exchange coupling is $J \approx 1$ meV, then the temperatures considered in Table \ref{tab:Thermal} correspond to $T \lesssim 10$ K.

\begin{table}[bh]
    \centering
    \caption{Quantum skyrmion order parameter, $Q_{\text{BL}}$, as a function of temperature $T$. (a) and (b) show the results in SkX1 at $K/J = 0.10$ and $K/J = 0.50$ where the magnon gap is $\Delta/J \approx 0.59$ and $\Delta/J \approx 0.13$, respectively. (c) Results in SkX2 at $K/J = 0.53$ where the magnon gap is $\Delta/J \approx 0.11$. The remaining parameters are $D/J = 2.16, U/J = 0.35$ and $S = 1$.}
    \begin{tabular}{cD{.}{.}{5}D{.}{.}{5}D{.}{.}{5}D{.}{.}{5}D{.}{.}{5}D{.}{.}{3}}
        \hline\hline
        \multicolumn{7}{c}{(a) SkX1, $K/J = 0.10$, $\Delta/J \approx 0.59$}\\
        \hline
        $T/J$              & 0.0    & 0.2     & 0.4     & 0.6     & 0.8     & 1.0   \\
        $Q_{\text{BL}}/N'$ & 0.949  & 0.947   & 0.923   & 0.876   & 0.811   & 0.736 \\
        \hline
        \multicolumn{7}{c}{(b) SkX1, $K/J = 0.50$, $\Delta/J \approx 0.13$}\\
        \hline
        $T/J$              & 0.00    & 0.04     & 0.08     & 0.12     & 0.16     & 0.20   \\
        $Q_{\text{BL}}/N'$ & 0.897   & 0.896    & 0.893    & 0.885    & 0.875    & 0.863  \\
        \hline
        \multicolumn{7}{c}{(c) SkX2, $K/J = 0.53$, $\Delta/J \approx 0.11$}\\
        \hline
        $T/J$               & 0.00     & 0.06    & 0.12     & 0.18     & 0.24 &  \\
        $Q_{\text{BL}}/N'$  & 0.99971  & 0.99970 & 0.99967  & 0.99961  & 0.99953 & \\
        \hline\hline
    \end{tabular}
    \label{tab:Thermal}
\end{table}

%----------------------------BIBLIOGRAPHY---------------------------------------
\bibliography{main.bbl}

\end{document}